\documentclass[twoside]{article}  
\usepackage{amsfonts}
\usepackage{amsbsy}
\usepackage{amsmath}
\newtheorem{Th}{Theorem}[section]
\newtheorem{Lemma}[Th]{Lemma}

\newtheorem{example}[Th]{Example}

\newtheorem{definition}[Th]{Definition}

\newtheorem{remark}[Th]{Remark}

\newcommand{\pf}{\paragraph{Proof}}
\newcommand{\pfend}{\par\vspace{2ex}\noindent}
\newcommand{\eind}{\hspace*{\fill}$\Box$\par\vspace{2ex}\noindent}
\newcommand{\ruimte}{\par\vspace{1ex}\noindent}

\newcommand{\sfour}{ \;\;\;\; }
\newcommand{\es}{x}

\newcommand{\B}{{\cal B}}
\newcommand{\Zpr}{{\mathbb Z}_{p^r}}
\newcommand{\Z}{\mathbb{Z}}
\newcommand{\F}{\mathbb{F}}
\newcommand{\dubbel}[1]{{\mathbb #1}}

\newcommand{\A}{{\cal A}}
\newcommand{\N}{\mathbb{N}}
\newcommand{\RR}{\mathcal{R}}
\newcommand{\beq}{\begin{equation}}
\newcommand{\eeq}{\end{equation}}
\newcommand{\bmat}{\left[ \begin{array}}
\newcommand{\emat}{\end{array} \right]}
\newcommand{\twee}[2]{\left[ #1 \sfour #2 \right]}

\newcommand{\AND}{\;\mbox{and }}

\newcommand{\DEG}{\mathrm{deg }\;}

\newcommand{\ORD}{\mathrm{ord }\;}

\newcommand{\FOR}{\sfour\mbox{for }}

\newcommand{\DIM}{\mathrm{dim }\;}
\newcommand{\PDIM}{p\mathrm{-dim }\;}
\newcommand{\PSPAN}{p\mathrm{-span }\;}
\newcommand{\SPAN}{\;\mathrm{span }\;}
\newcommand{\ellL}{L}
\newcommand{\bbold}{\mbox{\boldmath $b$}}
\newcommand{\wbold}{\mbox{\boldmath $w$}}
\DeclareMathOperator{\e}{e}
\DeclareMathOperator{\TOP}{TOP}
\DeclareMathOperator{\POT}{POT}
\DeclareMathOperator{\lt}{lt}
\DeclareMathOperator{\lc}{lc}
\DeclareMathOperator{\ord}{ord}

\DeclareMathOperator{\lm}{lm}
\DeclareMathOperator{\lpos}{lpos}
\DeclareMathOperator{\LL}{L}

\title{\LARGE \bf
Minimal Gr\"obner bases and the predictable leading monomial property}

\author{M.\ Kuijper\footnote{M.\ Kuijper is with the Department of Electrical and Electronic Engineering, University of Melbourne, VIC 3010, Australia {\tt\small m.kuijper@ee.unimelb.edu.au}} and K.\ Schindelar\footnote{K.\ Schindelar is with the Lehrstuhl D f\"ur Mathematik, RWTH Aachen University Templergraben 64, 52062 Aachen, Germany {\tt\small Kristina.Schindelar@math.rwth-aachen.de}}
\thanks{This research is partially supported by the Australian Research Council (ARC) and the Deutscher Akademischer Austausch Dienst (DAAD)
and co-financed by the Deutsche Forschungsgemeinschaft (DFG)}}

\begin{document}
 
\maketitle
\begin{abstract}
We focus on Gr\"obner bases for modules of univariate polynomial
vectors over a ring. We identify a useful property,
the ``predictable leading monomial (PLM) property'' that is shared by minimal Gr\"{o}bner bases of modules in $\F[x]^q$, no
matter what positional term order is used. The PLM
property is useful in a range of applications and can be seen as a strengthening of
the wellknown predictable degree property (= row reducedness), a terminology
introduced by Forney in the 70's. Because of the
presence of zero divisors, minimal Gr\"{o}bner bases over a finite
ring of the type $\Zpr$ (where $p$ is a prime integer and $r$ is an
integer $>1$) do not necessarily have the PLM property. In this paper we show how to derive,
from an ordered minimal Gr\"{o}bner basis, a so-called "minimal
Gr\"{o}bner $p$-basis" that does have a PLM property. We demonstrate that minimal Gr\"obner $p$-bases lend themselves
particularly well to derive minimal realization parametrizations over
$\Zpr$. Applications are in coding and sequences over $\Zpr$.
\end{abstract}
\section{Introduction}
Gr\"obner bases have proved useful tools for dealing with polynomial
vectors, with applications particularly in multidimensional
system theory. These applications range from controller design to minimal
realization of linear systems over fields.
Fundamental linear algebraic results on polynomial matrices over
fields can be
elegantly achieved via the theory of Gr\"obner
bases~\cite{AdamsLoustaunau,buchberger01}. In 
particular, the wellknown Smith-McMillan form
as well as the Wiener-Hopf form (``row reducedness'') can be achieved. Using the
theory of Gr\"obner bases these 
are two sides of the same coin, obtained by choosing a different
positional term order~\cite{Hermo,SchindelarL09}.
\ruimte
In this paper we focus on Gr\"obner bases for modules of polynomial
univariate vector polynomials, i.e., elements of $\RR [x]^q$, where $q$ is an integer
$\geq 1$. In the field case $\RR = \F$ all modules are free and a minimal Gr\"obner basis of a module $M$ is a basis in a linear algebraic sense. It is known that, for certain types of positional term orders, minimal Gr\"obner bases of modules in $\F[x]^q$ are extremely useful for a range of minimal interpolation-type problems. In this paper we attribute this usefulness to a property that we call the "predictable leading monomial (PLM) property". This property is shared by minimal Gr\"obner bases in $\F[x]^q$, irrespective of the particular positional term order that is used.
In the case that $\RR$ is a ring it may happen that a minimal Gr\"obner basis of a module $M$ in $\RR[x]^q$ is not a basis; this may happen even when $M$ is a free module.  
\ruimte
Motivated by coding applications, we consider modules over the finite ring $\Zpr$, where $p$ is a
prime integer and $r$ is a positive integer. It was shown in~\cite{kuijperPP07} that any module $M$ in $\Zpr[x]^q$ has a particular type of basis, called "reduced $p$-basis"; \cite{kuijperPP07} gives a constructive procedure that starts from any set of polynomial vectors that generate $M$. Using Gr\"obner theory, in this paper we derive an expression for such a reduced $p$-basis in terms of a minimal Gr\"obner basis with respect to the TOP (Term Over Position) order. Our result is valid for any choice of positional term order, not just TOP. We show that our $p$-basis (which we call "minimal Gr\"obner $p$-basis") has a PLM property with respect to the chosen positional term order. This PLM property is stronger than the "$p$-predictable degree property" from~\cite{kuijperPP07} and makes minimal Gr\"obner $p$-bases ideally suitable for minimal interpolation-type problems, as illustrated in subsection~\ref{subsec_appl}.
\ruimte
There are several advantages to the Gr\"obner approach. Firstly,
it offers flexibility through the choice of positional term order. This makes it possible to derive several analogous results
at once. Secondly, the approach offers 
scope for extension to other areas where Gr\"obner bases
are a standard tool, such as multidimensional systems.
Finally, a third
advantage of the Gr\"obner approach is that computational packages are
available to compute minimal Gr\"obner bases, such as
the {\sc Singular} computer algebra system~\cite{Singular}. 
A preliminary version of this paper is~\cite{kuijperScdc09}.
\section{Preliminaries on Gr\"obner bases}
In this section we present basic notions from Gr\"obner theory and summarize wellknown results. 
Most textbooks introduce Gr\"obner theory in the context of multivariate polynomials, that is, elements of $\RR[x_1 , \ldots
  , x_n]$, where $\RR$ is a ring. Instead, here we focus on univariate {\em vector}
polynomials, i.e., elements of $\RR [x]^q$, where $q$ is an integer
$\geq 1$. It is wellknown~\cite[Ex. 4.1.14]{AdamsLoustaunau} that multivariate Gr\"obner theory can be translated into univariate Gr\"obner theory for $\RR [x]^q$ by using positional monomial orders, such as TOP (Term Over Position) and
POT (Position Over Term), defined below, see also~\cite[sect.\
  3.5]{AdamsLoustaunau},~\cite[sect.\ 10.4]{beckerW93},~\cite[p.\ 89; p.\ 104]{oberst90} and
the recent survey paper~\cite{linXB08}. We focus on 
properties of Gr\"obner bases rather than construction of
Gr\"obner bases. For more details on construction the reader is referred
to~\cite{AdamsLoustaunau,Pauer,BrickensteinDreyerGreuelWedlerWienand}.
\ruimte
Throughout this paper $\RR$ is assumed to be a noetherian ring, i.e., all its ideals are finitely generated. 
\ruimte
The concepts of ``degree'' and ``leading coefficient'' for polynomials
in $\RR[\es]$ are extended to
polynomial {\em vectors} in $\RR [\es]^q$, as follows.
Let $\e_1, \dots, \e_q$ denote the unit vectors in $\RR^q$. 
The elements $x^\alpha \e_i$ with $i \in \{1, \dots, q\}$ and 
$\alpha \in \N_0$ are called \textbf{monomials}.
Several positional term orders can be defined on these monomials;
we recall the following two monomial orders (adopting the terminology of~\cite{AdamsLoustaunau}):
\begin{itemize}
 \item[$\bullet$] The {\bf Term Over Position (TOP)} order, defined
 as 
$$ x^{\alpha} \e_i < x^{\beta} \e_j \;\; :\Leftrightarrow \;\; \alpha < \beta
\mbox{ or } ( \alpha=\beta \mbox{ and } i>j ). $$ 
 \item[$\bullet$] The {\bf Position Over Term (POT)} order, defined
 as 
$$ x^{\alpha} \e_i < x^{\beta} \e_j \;\; :\Leftrightarrow \;\; i>j \mbox{ or } ( i=j \mbox{ and }  \alpha < \beta ) .$$  
\end{itemize}
Weighted and/or reflected versions of these orders are also possible as in~\cite{fitzpatrick95}. Clearly, whatever order is chosen, every nonzero element $f \in \RR[x]^q$ can be written
uniquely as
$$ f=\sum_{i=1}^L c_i X_i, $$
where $L \in \N$, the $c_i$'s are nonzero
elements of $\RR$ for $i=1 , \ldots , L$ and the polynomial vectors $X_1, \ldots , X_L$ are
monomials, ordered as $X_1 > \dots > X_L$. Using the terminology of~\cite{AdamsLoustaunau} 
we define
\begin{enumerate}
\item[$\bullet$] $\lm(f):=X_1$ as the \textbf{leading monomial} of $f$
\item[$\bullet$] $\lt(f):=c_1X_1 $ as the \textbf{leading term} of $f$
\item[$\bullet$] $\lc(f):=c_1$ as the \textbf{leading coefficient} of $f$
\end{enumerate}
Writing $X_1=x^{\alpha_1} \e_{i_1}$, where $\alpha_1 \in \N_0$ and $i_1
\in \{1,\ldots , q \}$, we define
\begin{enumerate}
 \item[$\bullet$] $\lpos(f):=i_1$ as the \textbf{leading position} of $f$
 \item[$\bullet$] $\deg(f):=\alpha_1$ as the \textbf{degree} of $f$.
\end{enumerate}
Note that for the TOP order the degree of $f$
equals the highest degree of its nonzero components in $\RR[\es]$,
whereas for the POT order it equals the degree of the first nonzero
component. Further, for the POT order the leading position of $f$
is the position of the first nonzero
component, whereas for the TOP order the leading position of $f$ is
the position of the first nonzero component of highest degree. 
\ruimte
Below we denote the
submodule generated by a polynomial vector $f$ by $\langle
f \rangle$. There are several ways to define Gr\"obner bases, here we
adopt the definition of~\cite{AdamsLoustaunau} which requires us to
first define the concept of ``leading term submodule'':
\begin{definition}
Let $F$ be a subset of $\RR [x]^q$. Then the submodule $\LL(F)$, defined as
$$\LL(F):= \langle \lt(f) \ | \ f \in F \rangle$$
is called the \textbf{leading term submodule} of $F$.
\end{definition}
\begin{definition}\label{def_grob}
Let $M \subseteq \RR[x]^q$ be a module and $G \subseteq M$. Then
$G$ is called a \textbf{Gr\"obner basis} of $M$ if $$ \LL(G) = \LL(M).$$
\end{definition}
It is
wellknown~\cite[Corollary 4.1.17 and Ex. 4.1.14]{AdamsLoustaunau} that a finite Gr\"obner basis
exists for any module in $\RR[x]^q$. In general, it can be shown that
a Gr\"obner basis $G$ of a module $M$ generates $M$, see also Lemma~\ref{GBsubGen} below.
The following lemma follows immediately from Definition~\ref{def_grob}.
\begin{Lemma}\label{leadVector}
Let $M$ be a submodule of $\RR[x]^q$ with Gr\"obner basis 
$G=\{g_1, \dots, g_m\}$ and let $0 \neq f \in M$. Then there exists a
subset $\{g_{j_1}, \cdots, g_{j_s}\}$ of $G$ and
$\alpha_1, \dots, \alpha_s \in \N_0$ and $c_1, \dots, c_s \in \RR$,
such that
\begin{itemize}
\item $\lm(f)=x^{\alpha_i}\lm(g_{j_i})$ for $i = 1, \ldots , s$ \AND 
\item $\lt(f)=c_1 x^{\alpha_1}\lt(g_{j_1}) + \dots + c_s
  x^{\alpha_s}\lt(g_{j_s})$. 
\end{itemize}
\end{Lemma}
Note that the $g_{j_i}$'s of the above lemma all satisfy $\lpos
(g_{j_i}) = \lpos (f)$ and $\lm (g_{j_i}) \leq \lm (f)$. The above
lemma inspires the next definition.
\begin{definition}\label{DefRed}{\rm (\cite[Def.\ 4.1.1]{AdamsLoustaunau}) }
 Let $0 \neq f \in \RR[x]^q$ and let $F=\{f_1, \dots, f_s\}$ be a set
 of nonzero elements of $\RR[x]^q$. Let
 $\alpha_1, \dots, \alpha_s \in \N_0$ and let $c_1, \dots, c_s$ be
 elements of $\RR$ such that
\begin{enumerate}
\item $\lm(f)=x^{\alpha_i}\lm(f_i)$ for $i = 1,\ldots , s$ and
\item $\lt(f)=c_1 x^{\alpha_1}\lt(f_1) + \dots + c_s x^{\alpha_s}\lt(f_s)$.
\end{enumerate}
Define 
\[
h := f- (c_1x^{\alpha_1} f_1 + \dots + c_sx^{\alpha_s} f_s) .
\]
Then we say that $f$ \textbf{reduces} to $h$ modulo $F$ and we write 
$$ f \xrightarrow{F} h . $$
If $f$ cannot be reduced modulo $F$, we say that $f$ is
\textbf{minimal} with respect to $F$. 
\end{definition}
\begin{Lemma}\label{lemma_smaller}{\rm (\cite[Lemma 4.1.3]{AdamsLoustaunau}) }
Let $f$, $h$ and $F$ be as in the above definition. If $f
\xrightarrow{F} h$ then $h=0$ or $\lm(h)<\lm(f)$. 
\end{Lemma}
The next lemma is an immediate corollary of Lemma~\ref{lemma_smaller} that will prove useful in the sequel.
\begin{Lemma}\label{GBsubGen}
Let $M$ be a submodule of $\RR[x]^q$ with Gr\"obner basis 
$G$ and let $0 \neq f \in M$. Then 
\[ f \in \langle g \in G \ | \ \lm(g) \leq \lm(f) \rangle. \] 
\end{Lemma}
\begin{definition}{\rm (\cite{AdamsLoustaunau})}
 A Gr\"obner basis $G$ is called \textbf{minimal} if all its elements $g$ are minimal with respect to $G\backslash\{g\}$.
\end{definition}
It is known~\cite[Exercises 4.1.9 \& 4.1.14]{AdamsLoustaunau} that a minimal Gr\"obner basis
exists for any module in $\RR[x]^q$ and that it
has the following convenient property:
\begin{Lemma}\label{GB-ordering}
Let $G=\{g_1, \dots, g_m\}$ be a minimal Gr\"obner basis. Then
$\lm(g_i) \neq \lm(g_j)$ for all $i,j \in \{ 1, \ldots , m\}$.
\end{Lemma}
\section{The field case}\label{sec_field}
In this section we limit our attention to the case that $\RR$ is a
field. It is wellknown that Gr\"obner bases are useful for various applications over fields, including univariate applications. In this section we attribute this usefulness to a particular property of minimal Gr\"obner bases.
We introduce the following terminology.
\begin{definition}\label{def_plm_field}
Let $M$ be a submodule of $\RR[x]^q$ and let $F
=\{f_1, \dots, f_s\}$ be a nonempty subset of $M$. Then $F$ has the \textbf{Predictable Leading Monomial (PLM) property} if for any $0 \neq f \in M$, written as
\beq
f=a_1f_1 + \dots + a_sf_s ,\label{eq_lin_comb_f}
\eeq
where $a_1, \dots, a_s \in \RR[x]$, we have
\beq
\lm(f)=\max_{1 \leq i \leq s; a_i \neq 0} ( \lm(a_i) \lm(f_i)) .\label{eq_plm}
\eeq
\end{definition}
\ruimte
In the Gr\"obner literature usually a weaker property than the
above PLM property is presented, namely: for any $f$ from a module $M$, generated by $f_1 , \ldots , f_s$, there {\em
  exist} $a_1, \ldots , a_s \in \RR[x]$ such
that~(\ref{eq_lin_comb_f}) and~(\ref{eq_plm}) hold, see~\cite[Thm
  1.9.1]{AdamsLoustaunau}. In the field case this is clearly
equivalent to the PLM property; for this reason the next theorem
merely reformulates a wellknown result.
\begin{Th}\label{thm_main_field}
Let $\RR$ be a field. Let $M$ be a submodule of $\RR[x]^q$ with
minimal Gr\"obner basis $G$. Then $G$ has the
Predictable Leading Monomial (PLM) property. In particular, $G$ is a
basis of $M$.
\end{Th}
\pf
{\rm Write $G = \{ g_1 , \ldots , g_m \}$. Since $G$ is minimal we may
assume, by Lemma~\ref{GB-ordering}, that $\lm(g_1) > \lm(g_2) > \cdots > \lm(g_m)$. Let $f = a_1 g_1 + \cdots +
a_m g_m $. For simplicity of notation we assume that $a_i$ is nonzero for $1 \leq
i \leq m$. Since $\RR$ is a field we have that $\lpos (a_i g_i ) =
\lpos (g_i )$ for $1 \leq i \leq m$. Also, all leading positions of the $g_i$'s
are distinct, otherwise we can reduce. As a result, all leading
monomials of the $a_i g_i$'s are distinct. Thus there exist $j_1, \ldots , j_m$, such that
\[
\lm(a_{j_1} g_{j_1}) > \lm ( a_{j_2} g_{j_2} )> \cdots > \lm(a_{j_m} g_{j_m} ).
\]
It follows that 
\[
\lm (f) = \lm (a_{j_1} g_{j_1} ) = \lm (a_{j_1}) \lm ( g_{j_1}) =\max_{1 \leq i \leq m} ( \lm(a_i) \lm(g_i)),
\]
which proves the PLM property. Finally, to prove that $G$ is a basis
of $M$, first observe that $G$ generates $M$ by
Lemma~\ref{GBsubGen}. Also, it follows immediately from the PLM
property that any nontrivial linear combination of vectors from $G$
has to be nonzero. We conclude that $G$ is a basis of $M$.}
\eind\pfend
Note that the PLM property is a strenghtening of the
well established predictable degree property from~\cite{forney70,forney75},
since it involves not only degree information but also leading
position information. Also, Theorem~\ref{thm_main_field} holds irrespective of the monomial order that is used. Of course, in the field case where all modules are free, the number $m$ of elements in a minimal Gr\"obner basis equals the dimension of $M$.
The next example demonstrates the usefulness of the PLM property, see also~\cite{fitzpatrick95}.
\begin{example}\label{example_lrr}: {\bf Using minimal Gr\"obner bases for parametrization of all shortest linear recurrence relations}
\newline
Consider the sequence $S_0 , S_1 , S_2, S_3, S_4 = 1,4,3,3,2$ over the field
$\Z_5$. A polynomial $d(x)$, written as
$d(x)=x^\ellL + d_{\ellL -1} x^{\ellL -1} + \cdots + d_1 x + d_0$,
is called a linear recurrence
  relation of length $\ellL$ for $S_0 , S_1 , S_2, S_3, S_4$ if
\beq
S_{\ellL + j} + \sum_{i=1}^\ellL d_{\ellL -i} S_{\ellL +j-i} = 0
\;\; \FOR j=0 , \ldots , 5-\ellL -1 .\label{eq_recur}
\eeq
Consider the polynomial $S(x):= S_0 x^5 + S_1 
x^4 + S_2 x^3 +S_3 x^2 + S_4 x $ and the module $M$ spanned by $\bmat{c@{\hspace{2em}}c} 1   & -S(x)\emat$ and $\bmat{c@{\hspace{2em}}c} 0   & x^6\emat$. Clearly, any minimal Gr\"obner basis for $M$ must consist of 2 vectors and exactly one of these 
vectors has leading position $1$. In fact, {\sc Singular} computes a
minimal TOP Gr\"obner basis $G = \{ g_1 , g_2 \}$ for $M$, with $g_1
(x) = \bmat{c@{\hspace{1em}}c} 2x+2   & x^4-2x^3+x\emat$ and $g_2 (x) = \bmat{c@{\hspace{1em}}c} x^2-3x-1   & 4x^2-3x\emat$.
The PLM property of $G$ implies that the
vector of leading position $1$, i.e. $g_2$, yields a unique shortest linear
recurrence relation, namely $x^2-3x-1$. The reader is also referred to
the recent paper~\cite{leeS08} where Gr\"obner bases are employed for similar problems.
\end{example}
Theorem~\ref{thm_main_field} does not extend to the case that $\RR$ is a ring. At first sight this may seem obvious as there exist modules in $\RR[x]^q$ that are not free. Evidently, any minimal Gr\"obner basis for such a module is not a basis so certainly does not satisfy the PLM property. However, the situation is more subtle: the Gr\"obner basis of a free module in $\RR[x]^q$ is not necessarily a basis either, as we will illustrate in Example~\ref{example_lrr_ring}. In this paper we are interested in solving this 
difficulty for the special case that $\RR$ is a ring of the type
$\Z_{p^r}$. For this we make use of the special structure of $\Z_{p^r}$. 
\section{The ring case}
\subsection{Preliminaries on $\Zpr$}
A set that plays a fundamental role
throughout this paper is the set of ``digits", denoted by $\A_p =
\{0,1,\dots, p-1\} \subset \mathbb Z_{p^r}$. Recall that any
element $a \in \dubbel{Z}_{p^r}$ can be written uniquely as
$a=\theta_0 +p \theta_1 + \cdots + p^{r-1} \theta_{r-1}$, where
$\theta_\ell \in \A_p$ for $\ell =0 , \ldots , r-1$ ({\em p-adic
expansion}). 
\ruimte
Next, adopting terminology from~\cite{vaziraniSR96}, an element $a$ in
$\Z_{p^r}$ is said to have \textbf{order} $k$ if the additive
subgroup generated by $a$ has $p^k$ elements. (Note
that~\cite{gorbatov04} and references therein use the terminology \textbf{norm} for $r-k$.)
Elements of order $r$ are called {\bf units}.
Thus the elements $1, p, p^2, \dots, p^{r-1}$ have orders $r, r-1,
r-2, \dots, 1$, respectively. Let us now choose a monomial order for
polynomial vectors in $Z_{p^r}[x]^q$. Given this
monomial order,
we now extend the above notion of ''order`` 
to polynomial vectors as follows.
\begin{definition}\label{deforderpol}
The \textbf{order} of a nonzero polynomial vector $f \in \mathbb
Z_{p^r}[x]^q$, is defined as the order of $\lc (f)$, denoted as $\ORD (f)$.
\end{definition}
To deal with the zero divisors occurring in $\Z_{p^r}$
it is useful to use notions of ``$p$-linear dependence'' and ``$p$-generator sequence'', first introduced for modules in $\Zpr^q$ in~\cite{vaziraniSR96}. These notions are based on the $p$-adic expansion property of $\Zpr$, which expresses a type of linear
independence among the elements $1$, $p$, ..., $p^{r-1}$. The notions
presented below are for {\em polynomial} vectors; they are extensions of~\cite{vaziraniSR96}, first presented in~\cite{kuijperPP07}.
\begin{definition} {\rm (\cite{kuijperPP07})}
Let $\{v_1, \dots, v_N \} \subset \mathbb Z_{p^r}[x]^q$. A
{\bf $\boldsymbol{p}$-linear combination} of $v_1, \dots, v_N$ is a
vector $\displaystyle \sum_{j=1}^N a_j v_j,$ where $a_j
\in \mathbb Z_{p^r}[x]$ is a polynomial with coefficients in
$\A_p$ for $j=1, \dots, N$. Furthermore, the set of all $p$-linear
combinations of $v_1, \dots, v_N$ is denoted by
{\bf $\boldsymbol{p}$-span}$(v_1, \dots, v_N)$, whereas the set of all linear
combinations of $v_1, \dots, v_N$ with coefficients in
$\mathbb Z_{p^r}[x]$ is denoted by $\SPAN (v_1, \dots,v_N)$.
\end{definition}

\begin{definition}\label{def_pgen} {\rm (\cite{kuijperPP07})}
An ordered sequence $(v_1, \dots, v_N)$ of vectors in
$\mathbb Z_{p^r}[x]^q$ is said to be a {\bf $\boldsymbol{p}$-generator
sequence} if $p \, v_N=0$ and $p \, v_i$ is a $p$-linear
combination of $v_{i+1}, \dots, v_N$ for $i=1, \dots, N-1$.
\end{definition}
\begin{Th}{\rm (\cite{kuijperPP07})} Let $v_1, \dots, v_N \in \mathbb Z_{p^r}[x]^q$. If
$(v_1, \dots, v_N)$ is a $p$-generator sequence then
\[
\PSPAN (v_1, \dots, v_N)= \SPAN (v_1, \dots, v_N) .
\]
In particular, $\PSPAN (v_1, \dots, v_N)$ is a submodule of
$\mathbb Z_{p^r}[x]^q$.
\end{Th}
All submodules of $\mathbb Z_{p^r}[x]^q$ can be written as the
$p$-span of a $p$-generator sequence. In fact, if $M= \SPAN
(g_1, \dots, g_m)$ then $M$ is the $p$-span of the
$p$-generator sequence
\[
(g_1,p g_1, \dots, p^{r-1}g_1,g_2,p g_2, \dots, p^{r-1}g_2, \dots, g_m, p g_m, \dots, p^{r-1}g_m) .
\]
\begin{definition} {\rm (\cite{kuijperPP07})}
The vectors $v_1, \dots, v_N \in \mathbb Z_{p^r}[x]^q$ are
said to be {\bf $\boldsymbol{p}$-linearly independent} if the only $p$-linear
combination of $v_1, \dots, v_N$ that equals zero is the
trivial one.
\end{definition}

\begin{definition}\label{def_pbasis} {\rm (\cite{kuijperPP07,kuijperP_it09})}
Let $M$ be a submodule of $\dubbel{Z}_{p^r}[x]^q$, written as a
  $p$-span  of a $p$-generator sequence $(v_1 , \cdots
  , v_N )$. Then  $(v_1 , \cdots , v_N )$ is
  called a
  {\bf $\boldsymbol{p}$-basis} of $M$ if the vectors $v_1, \dots,
  v_N$ are $p$-linearly independent in $\dubbel{Z}_{p^r}[x]^q$. The
  number of elements of a $p$-basis is called the {\bf
    $\boldsymbol{p}$-dimension} of $M$, denoted as $\PDIM(M)$.
\end{definition}
The following definition adjusts the PLM property, introduced for the
field case in Definition~\ref{def_plm_field}, to the specific structure of $\Zpr$. 
\begin{definition}
Let $M$ be a submodule of $\Z_{p^r}[x]^q$ and let $F=\{f_1, \dots,
f_s\}$ be a nonempty subset of $M$. Then $F$ has the {\bf $\boldsymbol{p}$-Predictable
  Leading Monomial ($\boldsymbol{p}$-PLM) property} if for any $0 \neq f \in M$, written as
\beq
f=a_1f_1 + \dots + a_s f_s ,\label{eq_plin_comb_f}
\eeq
where $a_1, \dots, a_s \in \A_p[x]$, we have
\[
\lm(f)=\max_{1 \leq i \leq s; a_i \neq 0} ( \lm(a_i) \lm(f_i)) .
\]
\end{definition}
Note that in the above definition $a_i \in \A_p[x]$ rather than $a_i
\in \RR [x]$ as in Definition~\ref{def_plm_field}. 
Further note that multiplications and additions
in~(\ref{eq_plin_comb_f}) are over $\Zpr$; also observe that $\A_p[x]$
is not closed under addition, for example, in $\Z_4[x]$,
we have $x\in \A_2[x]$ but $x+x =2x \notin \A_2[x]$.
\subsection{Main result}
By Lemma~\ref{GB-ordering}, a minimal Gr\"obner basis $G =\{g_1, \dots, g_m\}$ has the convenient
property that its elements can be ordered so that $\lm(g_1) > \dots > \lm(g_m)$ since
their leading monomials are distinct. Unlike
the field case, a minimal Gr\"obner basis of a module in
$\Z_{p^r}[x]^q$ is, in general, {\em not} a basis. In
fact, the leading positions of its elements are not necessarily distinct. This may happen even when the module is free. We
have the following lemma.
\begin{Lemma}\label{order}
Let $M$ be a submodule of $\Z_{p^r}[x]^q$ with minimal Gr{\"o}bner
basis $G =\{g_1, \dots, g_m\}$, ordered so that $\lm(g_1) > \dots >
\lm(g_m)$. Let $j
< i$ be such that $\lpos (g_j ) = \lpos (g_i )$. Then $\deg g_j > \deg
g_i$ and $ \ORD(g_j) > \ORD(g_i)$. In particular, $m \leq qr$.
\end{Lemma}
\pf
{\rm Since $\lpos(g_j)=\lpos(g_i)$ and $\lm (g_j) > \lm (g_i)$ we must have that 
$\deg(g_j) > \deg(g_i)$, regardless of the monomial order that is used. It then follows that $\ORD(g_j) > \ORD(g_i)$, otherwise
$g_j$ could be reduced by $g_i$ and this would contradict the fact
that $G$ is a minimal Gr{\"o}bner basis. This proves the main result
of the lemma. Since there are only $r$
values of $\ORD(g_i)$ possible, it also follows that $m \leq qr$.}
\eind\pfend
As a result of the previous lemma we can define a sequence of ''order
differences`` as follows.
\begin{definition}\label{def_orderdiff}
Let $M$ be a submodule of $\Z_{p^r}[x]^q$ with minimal Gr\"obner basis
$G=\{g_1, \dots, g_m\}$ ordered so that $\lm(g_1) > \dots > \lm(g_m)$. For $1 \leq j \leq m$ define
$$ \beta_j := \ORD(g_j) - \ORD(g_i),$$
where $i$ is the smallest integer $> j$ with $\lpos(g_i)=\lpos(g_j)$.
If $i$ does not exist we define $ \beta_j := \ORD(g_j)$.
The sequence $(\beta_1, \dots, \beta_m) \in \N^m$ is called the \textbf{sequence of order differences} of $G$.
\end{definition}
The next theorem shows that the natural ordering of elements of a
minimal Gr\"obner basis yields a particular $p$-generator sequence. Note
that the theorem holds for any choice of monomial order. 
\begin{Th}\label{thm_main1}
Let $M$ be a submodule of $\Z_{p^r}[x]^q$ with minimal Gr\"obner basis
$G=\{g_1, \dots, g_m\}$, ordered so that $\lm(g_1) > \dots >
\lm(g_m)$. Let
 $(\beta_1, \dots, \beta_m)$ be the sequence of order differences of
 $G$ as per Definition~\ref{def_orderdiff}. Then
\beq
( g_1 , p g_1 , \cdots , p^{\beta_1 -1}g_1 , g_2 ,  p g_2 , \cdots ,
p^{\beta_2 -1}g_2 , \cdots, g_m, p g_m , \cdots , p^{\beta_m -1}g_m )
\label{eq_pbasis}
\eeq
is a $p$-generator sequence whose $p$-span equals $M$.
\end{Th}
\pf
{\rm We first prove that~(\ref{eq_pbasis}) satisfies
  Definition~\ref{def_pgen}. By definition $\beta_m = \ORD(g_m)$, so that 
\begin{eqnarray}\label{ltRed}
 \lm(p^{\beta_m}g_m)<\lm(g_m).
\end{eqnarray}
Suppose $p^{\beta_m}g_m \neq 0$, then according to Lemma~\ref{leadVector}
there exists $g_i \in G$ such that $\lm(g_i)\leq \lm(p^{\beta_m}g_m)$. But then~(\ref{ltRed})
implies that $\lm(g_i)<\lm(g_m)$ which contradicts $\lm (g_1) > \dots >
\lm (g_m )$. We conclude that
\beq
p^{\beta_m}g_m = 0 .\label{eq_last}
\eeq
To prove that~(\ref{eq_pbasis}) satisfies Definition~\ref{def_pgen} it
now obviously remains to prove that 
$p^{\beta_j}g_j$ is a $p$-linear combination of
\beq
g_{j+1},pg_{j+1},\ldots ,p^{\beta_{j+1}-1}g_{j+1},g_{j+2},pg_{j+2},\ldots,p^{\beta_{j+2}-1}g_{j+2},\ldots,g_m,\ldots, p^{\beta_m-1}g_m\label{eq_plin}
\eeq
for $1 \leq j \leq m-1$. For this, we first prove that $p^{\beta_j}g_j$ is a linear combination of $g_{j+1} ,g_{j+2},\ldots
  , g_m$. We distinguish two cases:
\newline\noindent
\underline{case I}
\begin{quote}
$\beta_j = \ord g_j$. Then
  $\lm(p^{\beta_j}g_j)<\lm(g_j)$, so that, by Lemma~\ref{GBsubGen},
  $p^{\beta_j}g_j$ is a linear combination of $g_{j+1} ,g_{j+2},\ldots
  , g_m$.
\end{quote}
\underline{case II}
\begin{quote}
$\beta_j < \ord g_j$, so that $\lm(p^{\beta_j}g_j) = \lm(g_j)$. By definition, there exists a
  smallest integer $i > j$ with $\lpos (g_i ) = \lpos (g_j )$ and
  $\beta_j = \ord (g_j ) - \ord (g_i )$. Observe that then $\ord
  (p^{\beta_j}g_j ) = \ord (g_i )$ and $\deg (p^{\beta_j}g_j ) = \deg
  (g_j ) > \deg (g_i )$ (use Lemma~\ref{order}), whereas $\lpos
  (p^{\beta_j}g_j ) = \lpos (g_j ) = \lpos (g_i )$. Thus we can find 
$a \in Z_{p^r}[x]$ such that $\lt(p^{\beta_j}g_j)=\lt(a g_i)$. As
a result, $\lm(p^{\beta_j}g_j - a g_i ) <\lm(p^{\beta_j}g_j) =
\lm(g_j)$. Consequently, by Lemma~\ref{GBsubGen},
  $p^{\beta_j}g_j - a g_i$ is a linear combination of $g_{j+1} ,g_{j+2},\ldots
  , g_m$. Since $i > j$ it follows that $p^{\beta_j}g_j$ is also a
  linear combination of $g_{j+1} ,g_{j+2},\ldots , g_m$.
\end{quote}
Thus for $1 \leq j \leq m-1$
\beq
p^{\beta_j}g_j \;\;\mbox{is a linear combination of}\;\; g_{j+1},\ldots  , g_m.\label{eq_lin} 
\eeq
Finally, we prove by induction that~(\ref{eq_plin}) holds for $1 \leq j
\leq m-1$. For $j=m-1$ this follows from~(\ref{eq_last}) and the fact
that $p^{\beta_{m-1}}g_{m-1}$ is a multiple of $g_m$
because of~(\ref{eq_lin}). Now suppose that~(\ref{eq_plin}) holds for $j=j_0
\in \{1, \ldots ,m-1 \}$. Consider the vector
$p^{\beta_{j_0 -1}}g_{j_0 -1}$. By~(\ref{eq_lin}) there exist $a_{j_0},
\ldots , a_m \in \Zpr[x]$ such that
\[
p^{\beta_{j_0 -1}}g_{j_0 -1} = a_{j_0}g_{j_0} + \cdots + a_m g_m .
\]
Now use the $p$-adic decomposition to write
\[
a_{j_0} = a^0_{j_0} + p a^1_{j_0} + \cdots + p^{r-1}a^{r-1}_{j_0} ,
\]
where $a^i_{j_0} \in \A_p [x]$ for $0 \leq i \leq r-1$. Repeatedly
using the induction hypothesis it follows that
\[
p^{\beta_{j_0 -1}}g_{j_0 -1} = a^0_{j_0}g_{j_0} + \mbox{$p$-linear
  combination of}\;\; g_{j_0 +1},\ldots  , g_m .
\]
This proves that~(\ref{eq_plin}) holds for $j= j_0 -1$, so that, by
induction, (\ref{eq_pbasis}) is a $p$-generator sequence.
\ruimte
To prove that its $p$-span equals $M$, we first note that, by
Lemma~\ref{GBsubGen}, any element of $M$ can be written as a linear
combination of $g_1, g_2, \ldots ,g_m$. Using a similar reasoning as
above this can be alternatively written as a $p$-linear combination of
the vectors in~(\ref{eq_plin}).}
\eind\pfend
The next lemma follows immediately from Definition~\ref{def_orderdiff}.
\begin{Lemma}\label{lemma_posord}
Let $M$ be a submodule of $\Z_{p^r}[x]^q$ with minimal Gr\"obner basis
$G=\{g_1, \dots, g_m\}$, ordered so that $\lm(g_1) > \dots >
\lm(g_m)$. Let
 $(\beta_1, \dots, \beta_m)$ be the sequence of order differences of
 $G$ as per Definition~\ref{def_orderdiff} and let $N=\beta_1 + \beta_2 +
\cdots + \beta_m$. Let $(v_1 , \ldots ,
 v_N )$ be the $p$-generator sequence given by~(\ref{eq_pbasis}).
Then for any $i,j \in \{1, \ldots , N\}$ with $i \neq j$ we have
\[
\lpos (v_i) = \lpos (v_j) \Rightarrow \ord (v_i) \neq \ord (v_j) .
\]
\end{Lemma}
The next theorem is the ring analogon of Theorem~\ref{thm_main_field}
and presents the main result of this section.
\begin{Th}\label{thm_main2}
Let $M$, $(\beta_1, \dots, \beta_m)$ and $\{ v_1 , \ldots , v_N \}$ be defined as in the previous lemma.
Then $\{ v_1 , \ldots , v_N \}$ has the $p$-PLM property.
\newline
In particular,
$(v_1 , \ldots , v_N )$ is a $p$-basis of $M$ so that
\[
N=\PDIM(M)=\beta_1 + \beta_2 +\cdots + \beta_m .
\]
\end{Th}
\pf
{\rm Let
\beq
f = a_1 v_1 + \cdots + a_N v_N \label{eq_flincomb}
\eeq
with $a_1, \dots, a_N \in \A_p[x]$. For simplicity of
notation we assume that $a_i$ is nonzero for $1 \leq i \leq N$. Let us first examine two special cases:
\newline\noindent
\underline{Special case I}
\begin{quote}
All $g_i$'s have distinct leading
  positions. Then the proof is analogous to the field case, i.e., the
  proof of Theorem~\ref{thm_main_field}.
\end{quote}
\underline{Special case II}
\begin{quote}
All $g_i$'s have the same leading
  position. Then all $v_i$'s also have the same leading position. By
  Lemma~\ref{lemma_posord} their orders are all different. Now observe
  that $\ord (a_i v_i ) = \ord (v_i )$ for $1 \leq i \leq N$ since
  $a_i \in \A_p[x]$. Thus all $a_i v_i$'s have different
  orders. In particular, all $a_i v_i$'s of largest degree have
  different orders, so that their leading coefficients add up to a
  nonzero element of $\Zpr$ (use the $p$-adic decomposition). This
  implies that the $p$-PLM property holds.
\end{quote}
Let us now consider the general case. By grouping together all vectors $a_i v_i$ of the same leading position we write
\[
f= f_1 + f_2 + \cdots + f_q ,
\]
where $f_i =0$ if position $i$ is not used in~(\ref{eq_flincomb}). As
in Special case II above it can be shown that $\lpos (f_i ) = i$
whenever $f_i \neq 0$. As a result, the nonzero $f_i$'s can be ordered
and it follows that
\beq
\lt (f) = \lt (f_j)\label{eq_fj}
\eeq
for some nonzero $f_j$ with $j \in \{1, \ldots , q\}$. Recall that
$f_j$ is defined as the sum of all vectors in the right hand side
of~(\ref{eq_flincomb}) that have leading position $j$. It now follows
from Special case II above that there exists $\ell \in \{1, \ldots ,
N\}$ such that $\lm (f_j ) = \lm (a_\ell ) \lm (v_\ell)$. As a result,
by equation~(\ref{eq_fj}), 
\beq
\lm (f) = \lm (a_\ell ) \lm (v_\ell) .\label{eq_flm}
\eeq
Evidently $\lm (f) \leq \max_{1 \leq i \leq N; a_i \neq 0} ( \lm(a_i)
\lm(v_i))$, so that~(\ref{eq_flm}) implies that equality holds. This
proves the $p$-PLM property. 
\ruimte
Finally, to prove that $(v_1 , \ldots , v_N)$ is a $p$-basis for $M$,
first observe that $p$-span $(v_1 , \ldots , v_N) = M$ by
Theorem~\ref{thm_main1}. Also, it follows immediately from the $p$-PLM
property that any nontrivial $p$-linear combination of vectors in
$\{v_1 , \ldots , v_N \}$ has to be nonzero. We conclude that $(v_1 ,
\ldots , v_N)$ is a $p$-basis of $M$, so that $N=\PDIM (M) = \beta_1 + \beta_2 +
\cdots + \beta_m$.}
\eind\pfend
\begin{remark}
We stress the difference between the above $p$-PLM property and the
property of a so-called ``strong Gr\"obner basis'' $G$ in the literature
(terminology from~\cite{nortonS03}) which states that for any $f \in
M$ there {\em exist} $a_1, \ldots , a_m \in \Zpr[x]$ such
that $\lm (f) = \max_{1 \leq i \leq m; a_i \neq 0} ( \lm(a_i) \lm(g_i))$,
see also~\cite[Thm 4.1.12]{AdamsLoustaunau}, as well as~\cite{nortonS03}
and~\cite[Th.\ 2.4.3]{byrneF01}.
In the terminology of~\cite{gorbatov04}, this is formulated as ``any $f
\in M$ possesses an $H$-presentation relative to $G$''. In the
Gr\"obner basis literature it seems to be generally accepted that
uniqueness of representation via Gr\"obner bases can not be obtained
for the ring case. However, in this paper we adopt the novel approach
of~\cite{kuijperPP07} of restricting coefficients to $\A_p[x]$ to
achieve the $p$-PLM property, which implies uniqueness of
representation.
\end{remark}
\begin{definition}\label{def_pGB}
Let $M$ be a submodule of $\Z_{p^r}[x]^q$ with minimal Gr\"obner basis
$G=\{g_1, \dots, g_m\}$, ordered so that $\lm(g_1) > \dots >
\lm(g_m)$. Let
 $(\beta_1, \dots, \beta_m)$ be the sequence of order differences of
 $G$ as per Definition~\ref{def_orderdiff}. Let $(v_1 , v_2 , \ldots ,
 v_N )$ be the $p$-generator sequence given by~(\ref{eq_pbasis}). Then
 $(v_1 , v_2 , \ldots , v_N )$ is called a {\bf minimal Gr\"obner
   $\boldsymbol{p}$-basis} for $M$.
\end{definition}
\begin{example}\label{example_lrr_ring}
 Let $M$ be a submodule of $\Z_{9}^2[x]$, given as $M=\SPAN \{ s_1 ,
 s_2 \}$, where $s_2 (x)= \twee{0}{x^6}$ and $s_1 (x) =
 \twee{1}{-S(x)}$ with $S(x):= x^5 + 4x^4 + 4 x^3 +7 x^2 + 7x $.
\begin{itemize}
 \item Using the $\TOP$ order:\\
a minimal Gr\"obner basis $G=\{g_1, \dots, g_4\}$ of $M$ is given by the rows of
$$
\left[ \begin{array}{cc}
8 &  \underline{x^5} +4x^4+4x^3+7x^2+7x\\
x+5 &    \underline{3x^4} +3x^2+x\\
\underline{x^2} +3x+2 & x^2+4x\\
\underline{3x} +6 &   3x\\
\end{array} \right].$$
Note that $M$ is a free module but $G$ is not a basis. The sequence of order differences $(\beta_1,
\beta_2,\beta_3,\beta_4)$ equals $(1,1,1,1)$. By
Theorem~\ref{thm_main2}, the sequence $(g_1,g_2,g_3,g_4)$ is a minimal
Gr\"obner $p$-basis for $M$ and therefore has the $p$-PLM property. Furthermore,
$\PDIM (M) = \beta_1 + \beta_2 +\beta_3 +\beta_4=4$.
 \item Using the $\POT$ order: \\
in this case the vectors $s_1$ and $s_2$ constitute a minimal Gr\"obner basis that happens to be a basis for $M$.
The sequence of order differences $(\beta_1 , \beta_2 )$ equals
$(2,2)$. According to Theorem~\ref{thm_main2}, the sequence
$(s_1,3s_1,s_2,3s_2)$ is a minimal Gr\"obner $p$-basis for $M$; it has the $p$-PLM
property. In fact, $\{s_1 ,s_2 \}$ has the PLM property as per Definition~\ref{def_plm_field}. Note that $\beta_1 +
\beta_2$ equals $4 = \PDIM (M) = 2\DIM (M)$, as expected.
\end{itemize}
\end{example}
Note that in this example the number of elements of the minimal POT Gr\"obner basis differs from the number of elements of the minimal TOP Gr\"obner basis, something that can't happen in the field case. However, the example clearly illustrates a corollary of Theorem~\ref{thm_main2}, namely that the sum of the $\beta_i$'s is an invariant of the module $M$, namely $N=\PDIM (M)$. Any minimal Gr\"obner $p$-basis of $M$ must consist of $N$ vectors, no matter which monomial order is used. 
Further, note that if the TOP order is used then a minimal Gr\"obner
$p$-basis is a ''reduced $p$-basis'' in the
terminology of~\cite{kuijperPP07}. Indeed, the $p$-PLM property clearly implies the $p$-predictable degree property of~\cite{kuijperPP07}.
Thus one of the applications where a minimal TOP Gr\"obner $p$-basis can be used is in convolutional coding over $\Zpr$: a minimal TOP Gr\"obner $p$-basis then serves as a minimal $p$-encoder of a convolutional code over $\Zpr$ in the terminology of~\cite{kuijperPcastle08, kuijperP_it09}. Applications for which the $p$-PLM property is particularly useful are parametrizations for minimal interpolation-type problems, as illustrated in the next subsection.
\subsection{An application over $\Zpr$}\label{subsec_appl} 
In the previous subsection we introduced the novel concept of ``minimal  Gr\"obner $p$-basis'' for modules over $\Zpr$. In this subsection we put this concept to work to
get a particularly transparent derivation of a parametrization of all
shortest linear recurrence relations of a finite sequence over $\Zpr$ that parallels the one
in~\cite{kuijperPmtns08}. In particular, we demonstrate the usefulness of the $p$-PLM property.
\ruimte
Consider the sequence $S_0 , S_1 , \ldots, S_{n-1}$ over $\Zpr$. We call a polynomial $f \in \Zpr[x]$, written as
$f(x)=f_\ellL x^\ellL + f_{\ellL -1} x^{\ellL -1} + \cdots + f_1 x + f_0$,
a {\bf linear recurrence
  relation} of length $\ellL$ for $S_0 ,  \ldots, S_{n-1}$ if $f_\ellL$ is a unit and
\beq
f_\ellL S_{\ellL + j} + \sum_{i=1}^\ellL f_{\ellL -i} S_{\ellL +j-i} = 0
\;\; \FOR j=0 , \ldots , n-\ellL -1 .\label{eq_recur_ring}
\eeq
As usual, we call the polynomial $f$ {\bf monic} if $f_\ellL = 1$. As in
Example~\ref{example_lrr}, define the polynomial $S(x)$ as 
\beq S(x ) := S_0 x^n + S_1 x^{n-1} +\dots +  S_{n-1} x ,
\label{eqS(x)}
\eeq
and consider $M=\SPAN \{ s_1 , s_2 \}$, where $s_1 (x)=\bmat{c@{\hspace{2em}}c} 1   &
-S(x)\emat$ and $s_2 (x)=\bmat{c@{\hspace{2em}}c} 0   & x^{n+1}\emat$. 
Obviously, $M$ is a free module for which $\{ s_1 , s_2 \}$ is a minimal POT Gr\"obner basis with $(\beta_1 , \beta_2 )=(r,r)$. Clearly, $\{ s_1 , s_2 \}$ is even a basis for $M$ and $\PDIM(M)=2r$. The theorem below parallels Theorem 15
of~\cite{kuijperPmtns08}, where Gr\"obner bases are not used; note that here no reordering of $p$-basis
vectors is required because the natural order of a minimal Gr\"obner
$p$-basis suffices. 
\begin{Th}\label{thm_lrr}
Let $S(x ) = S_0 x^n + S_1 x^{n-1} +\dots +  S_{n-1} x \in \Zpr[x]$ and
let
\[
M = \SPAN \{ \bmat{c@{\hspace{2em}}c} 1   &
-S(x)\emat , \bmat{c@{\hspace{2em}}c} 0   & x^{n+1}\emat \} .
\]
Let $(v_1 ,v_2 , \ldots , v_{2r})$ be a minimal TOP Gr\"obner
$p$-basis of $M$, with $v_i$ written as $v_i = \twee{d_i}{-h_i}\in 
\Zpr^2[x]$ for $i=1,\ldots , 2r$. 
Let $\ell \in \{ 1, \ldots , 2r \}$ be such that $\lpos (v_\ell) = 1$ and
$\ORD (v_\ell ) = r$. Then $d_\ell$ is a shortest linear recurrence
relation for the sequence $S_0 ,  \ldots, S_{n-1}$. Furthermore, a parametrization
of all shortest linear recurrence
relations for $S_0 ,  \ldots, S_{n-1}$ is given by
\beq
q_\ell d_\ell + \sum_{i>\ell} q_i d_i , \label{eq_para_Zpr}
\eeq
with $0\neq q_\ell \in \A_p$ and $q_i \in \A_p [x]$ with $\DEG q_i \leq \DEG v_\ell - \DEG v_i$ for $i=\ell+1 ,\ldots , 2r$. 
\end{Th}
\pf
{\rm We use a behavioral setup as in~\cite{kuijperPmtns08}. Consider the {\em partial impulse response behavior}
\[
\B := \SPAN \{\bbold ,
\sigma \bbold, \sigma^2 \bbold, , \ldots ,\sigma^n \bbold\}
,
\]
where $\bbold$ is defined as
\[
\bbold
=\left(\bmat{c} S_0 \\ 0 \emat , \bmat{c} S_1 \\ 0 \emat, \cdots , \bmat{c} S_{n-1}  \\ 0 \emat, \bmat{c} 0 \\ 1 \emat , \bmat{c} 0
\\ 0 \emat , \cdots \right) .
\]
It is easily verified that $M=\B^\perp$, that is, $M$ consists of all
annihilators of $\B$. As a result, $v_\ell$ is an annihilator of $\B$,
that is,
\[
\twee{d_\ell (\sigma   )}{-h_\ell (\sigma)} \wbold = 0
\] 
is a kernel representation whose behavior includes $\B$. Also $\DEG
h_\ell \leq \DEG d_\ell$. It then follows immediately that $d_\ell$ is
a linear recurrence relation for $S_0 , \ldots , S_{n-1}$. 
\newline
Next, suppose that a polynomial $d^\star \in \Z_{p^r}[x]$ is a
shortest linear recurrence relation for $S_0 ,\ldots , S_{n-1}$.
Then there exists a polynomial $h^\star \in \Z_{p^r}[x]$ of degree
$\leq \DEG d^\star$ such that $\twee{d^\star}{-h^\star}\in M$. Since
$(v_1 , v_2 , \ldots , v_{2r} )$ is a minimal Gr\"obner $p$-basis of
$M$ we can write $\twee{d^\star}{-h^\star}$ as a $p$-linear
combination of $v_1$, $v_2\ldots , v_{2r}$. Since $v_\ell$ is the
unique vector in this TOP Gr\"obner $p$-basis of leading position 1 and
order $r$, this $p$-linear combination {\em must} use
$v_\ell$. Because of the $p$-PLM property of $\{ v_1 , \ldots , v_N
\}$ (Theorem~\ref{thm_main2}), it follows that $\DEG d^\star \geq \DEG v_\ell$. This implies
that $v_\ell$ is a {\em shortest} linear recurrence relation for $S_0
, \ldots , S_{n-1}$. Moreover, it also follows from the $p$-PLM
property of $\{ v_1 , \ldots , v_N \}$ that the above $p$-linear
combination can {\em not} use $v_i$ for $i < \ell$. This proves the parametrization~(\ref{eq_para_Zpr}).}
\eind\pfend
\begin{example}~\label{ex_14477}
{\rm Consider the sequence $S_0 , S_1 , S_2, S_3, S_4 = 1,4,4,7,7$
  over the ring $\Z_9$. Let $M$ be the submodule of $\Z_{9}^2[x]$,
  defined as in Theorem~\ref{thm_lrr}. As shown
  in Example~\ref{example_lrr_ring}, a minimal TOP Gr\"obner $p$-basis
  $G=\{g_1, g_2 , g_3, g_4\}$ is given by the rows of
$$
\left[ \begin{array}{cc}
8 &  \underline{x^5} +4x^4+4x^3+7x^2+7x\\
x+5 &    \underline{3x^4} +3x^2+x\\
\underline{x^2} +3x+2 & x^2+4x\\
\underline{3x} +6 &   3x\\
\end{array} \right].$$
According to Theorem~\ref{thm_lrr}, $g_3$ gives a shortest linear
recurrence relation $x^2+3x+2$; a parametrization of all shortest linear
recurrence relations is given by
\[
(\Theta_1 (x^2+3x+2 ) + (\Theta_2 x + \Theta_3 )(3x+6 ) ,
\]
where $(\Theta_i \in \{ 0, 1,2 \}$ for $i=1,2,3;\;\Theta_1 \neq 0$. It
is easily seen that a parametrization of all {\em monic} shortest linear
recurrence relations is given by
\[
x^2+3x+2 + \Theta(3x+6 ) ,
\]
where $(\Theta \in \{ 0, 1,2 \}$, which concurs
with~\cite{kuijperPmtns08}. Note that this example illustrates that
non-uniqueness occurs despite the fact that the complexity is less
than $(n+1)/2$, a situation that does not occur in the field case.} 
\end{example}
\begin{example}~\label{ex_63156}
{\rm Consider the sequence $S_0 , S_1 , S_2, S_3, S_4 = 6,3,1,5,6$
  over the ring $\Z_9$, as in~\cite{ReedsS85}. The iterative
  algorithm of~\cite{ReedsS85} computes a shortest linear recurrence
  relation $x^3 +4x^2+7x+1$; note that no
parametrization is given in~\cite{ReedsS85}. Here we demonstrate how a
minimal TOP Gr\"obner $p$-basis can be used to derive a
parametrization. For this, let $M$ be the submodule of $\Z_{9}^2[x]$,
  defined as in Theorem~\ref{thm_lrr}. {\sc Singular} computes the
  minimal TOP Gr\"obner basis
  $G=\{g_1, g_2\}$ of $M$, where $g_1 = \twee{x^3 +4x^2+7x+4}{x^2 +
  3x}$ and $g_2 = \twee{6x^2 + 8}{x^3 +5x^2+6x}$. Note that, unlike Example~\ref{ex_14477}, $G$ is a basis for $M$.
According to Theorem~\ref{thm_main2}, the sequence
$(g_1,3g_1,g_2,3g_2)$ is a minimal TOP Gr\"obner $p$-basis for $M$. 
According to Theorem~\ref{thm_lrr}, $g_1$ gives a shortest linear
recurrence relation $x^3 +4x^2+7x+4$; a parametrization of all monic shortest linear
recurrence relations is given by
\[
x^3 +4x^2+7x+4 + \Theta_1 (6 x^2+8 ) + 6 \Theta_2  , 
\]
where $\Theta_i \in \{ 0, 1,2 \}$ for $i=1,2$. Thus for
$\Theta_1 =0$ and $\Theta_2=1$ we recover the
shortest linear recurrence
relation $x^3 +4x^2+7x+1$ from~\cite{ReedsS85}. In fact, $G$ has the PLM property as per Definition~\ref{def_plm_field} and the above parametrization can be rewritten as
\[
x^3 +4x^2+7x+4 + \Theta (6 x^2+8 )   , 
\]
where $\Theta \in \Z_9$. For $\Theta =3$ we recover the
shortest linear recurrence
relation $x^3 +4x^2+7x+1$ from~\cite{ReedsS85}.}
\end{example}
Note that both Example~\ref{ex_14477} and Example~\ref{ex_63156} are concerned with a free module. The two examples differ in the sense that $G$ is a basis in Example~\ref{ex_63156} but not a basis in Example~\ref{ex_14477}. This situation does not happen in the field case, where any minimal Gr\"obner basis of a module is a basis.
In Example~\ref{ex_14477} $G$ happens to be a $p$-basis that has the $p$-PLM property, whereas in Example~\ref{ex_63156} $G$ happens to be a basis that has the PLM property. In general, modules in $\Zpr[x]^q$ may have a minimal Gr\"obner basis that is neither a basis nor a $p$-basis. Our main result Theorem~\ref{thm_main2} shows how to construct a $p$-basis from $G$ that has the $p$-PLM property.
\section{Conclusions}
The main contributions of the paper are twofold. Firstly, we identified a particularly useful property, that we
labeled the ``Predictable Leading Monomial (PLM)'' property. A generating set of a module $M$ in $\RR[x]^q$ that has this property is necessarily a basis for $M$. For the case that $\RR$ is a field the PLM property is shared by minimal Gr\"obner bases of any module in $\RR[x]^q$. This is not necessarily the case when $\RR$ is a ring, even when the module is free. 
As our second main contribution, for any module $M$ in $\RR[x]^q$, we showed how to derive a particular set from a minimal Gr\"obner basis $G$ of $M$. We called this set a "minimal Gr\"obner $p$-basis" of $M$ and showed that it has a so-called "$p$-PLM property". The result is fairly trivial if $G$ happens to be a basis. However, the result is non-trivial in case $G$ is not a basis. We illustrated the latter with an example of a free module in $\Z_9 [x]^2$. To demonstrate the usefulness of the $p$-PLM property, we showed how to obtain
a parametrization of all
shortest linear recurrence relations of a finite sequence over $\Zpr$
from a minimal TOP Gr\"obner $p$-basis of a particular free module. Such parametrizations can be exploited
to decode beyond the minimum distance of Reed-Solomon codes, i.e., for list
decoding, see the recent paper~\cite{wu08}. Similarly, parametrizations
of interpolating solutions can be obtained for list decoding of Reed-Solomon codes.
\ruimte
One of the advantages of the Gr\"obner approach is its flexibility in the choice of monomial order.
This not only makes it possible to
derive several analogous results at once, but also makes it possible
to relate results obtained for different monomial orders. For example, in
the linear recurrence application we made use of the fact that a
minimal POT Gr\"obner $p$-basis of a module $M$ has the same number of elements as a
minimal TOP Gr\"obner $p$-basis of $M$. A possible topic of future
research is a Smith-McMillan like canonical form for polynomial matrices over
$\Zpr$. This is motivated by issues concerning catastrophicity of
convolutional codes over $\Zpr$, see~\cite{kuijperP_it09}.
\ruimte
The approach lends itself well to generalization to the multivariate case, see also~\cite {gorbatov04} and
references therein. This is another possible topic of future research.
\section*{Acknowledgment}
We thank one of the anonymous reviewers for detailed and helpful comments that have improved the clarity of the paper.


\end{document}